\documentclass[conference]{IEEEtran}
\IEEEoverridecommandlockouts
\usepackage{graphicx}
\usepackage{float}
\usepackage{filecontents}
\usepackage[noadjust]{cite}
\usepackage{verbatim}
\usepackage[nodisplayskipstretch]{setspace}
\usepackage{xcolor,colortbl}
\usepackage[]{algorithm2e}
\makeatletter
\ifCLASSINFOpdf
\else
\fi
\begin{document}
%
\title{Adaptive Data Collection Mechanisms for Smart Monitoring of Distribution Grids}

\author{\IEEEauthorblockN{Mohammed S. Kemal}
	\IEEEauthorblockA{Department of Electronic Systems\\
		Aalborg University\\
		Email: seifu@es.aau.dk}
	\and
	\IEEEauthorblockN{Rasmus L. Olsen}
	\IEEEauthorblockA{Department of Electronic Systems\\
		Aalborg University\\
		Email: rlo@es.aau.dk }

	
}

\maketitle

\begin{abstract}

Smart Grid systems not only transport electric energy but also information which will be active part of the electricity supply system. This has led to the introduction of intelligent components on all layers of the electrical grid in power generation, transmission, distribution and consumption units. For electric distribution systems, Information from Smart Meters can be utilized to monitor and control the state of the grid. Hence, it is indeed inherent that data from Smart Meters should be collected in a resilient, reliable, secure and timely manner fulfilling all the communication requirements and standards. This paper presents a proposal for smart data collection mechanisms to monitor electrical grids with adaptive smart metering infrastructures. A general overview of a platform is given for testing, evaluating and implementing mechanisms to adapt Smart Meter data aggregation. Three main aspects of adaptiveness of the system are studied, adaptiveness to smart metering application needs, adaptiveness to changing communication network dynamics and adaptiveness to security attacks. Execution of tests will be conducted in real field experimental set-up and in an advanced hardware in the loop test-bed with power and communication co-simulation for validation purposes.


\end{abstract}

\begin{IEEEkeywords}
Smart Grid, Adaptive Data Collection, AMI, QOS, Information Quality Monitoring
\end{IEEEkeywords}

\IEEEpeerreviewmaketitle

\section{Introduction}
   The current distributed, bi-directional power flow with increased integration of renewable sources has began to transform the traditional centralized one-directional chain of energy production, transmission and consumption to a Smart Grid. In this regard, Smart Meters play a major role, by giving suppliers access to accurate data used for billing, collecting end user informations, and establishing a two-way communication on the grid \cite{Smart_Meter}. The information can further be utilized to monitor the state of the grid. For example, the massive deployment of grid connected  PV systems and heat pumps  based on power electronic converters is introducing added operational challenges on Danish Distribution System Operators (DSO) \cite{REMOTEGRID}. The conventional approach used by DSOs to monitor the grid is pushed to levels where the system reaches its limits. To resolve this challenge,  DSOs can make use of Advanced Smart Metering infrastructures which are widely deployed in Denmark on low voltage grid to monitor and control the state of the grid.

	Modern Smart Metering infrastructure encompasses intelligent meters, data concentrators, central system units and different control centers defining a new smart metering architecture \cite{Smart_Meter_2} (see Figure 1). It is apparent that significant improvements on ICT has opened the door for resolving restrictions on near real time monitoring applications due to communication constraints. Modern communication technologies are at the heart of Smart Metering systems to ensure efficient information access mechanisms between Smart Meters, Concentrators and Head-End System \cite{On_Line}.

	\graphicspath{ {figures/} }
	\begin{figure}[H]
			\centering
		\includegraphics[width=4in,height=2.5in]{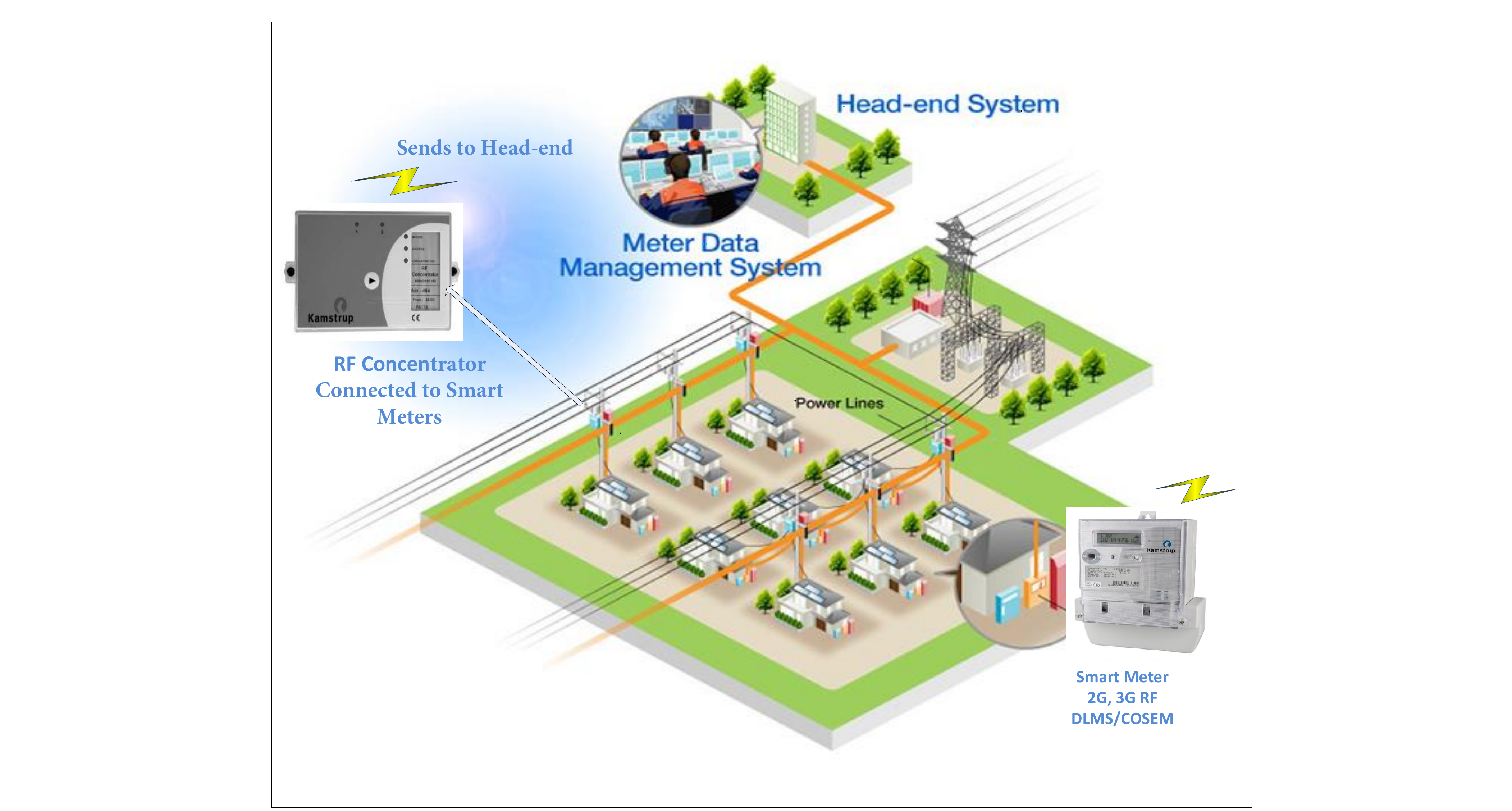}
		\label{AMI}
			\caption{Example of Smart Metering Infrastructure with Wireless 2G/3G Smart Meters, Kamstrup Concentrators and Head-end system for Metering Data Management Applications \cite{Kamstrup}  } 
	\end{figure}

	In this paper, a proposal for developing, testing and implementing mechanisms to adapt communication access schemes and protocols for Smart Metering infrastructure is presented. Focus is given for adaptiveness to Smart Metering application needs, to changing communication network dynamics and to security attacks. The proposed platform will be used to compare and contrast communication technologies, to develop and test adaptation mechanisms to ensure reliable transfer of information for data aggregation and to integrate, test standard Industrial Smart Meters, Concentrators and protocols.




 

\section{Related Works}

	Initially, the communication protocols used for Smart Metering implementations were not standardized but mostly proprietary owned. But, in recent years, various initiatives have been proposed to standardize and give interconnection assurance between Smart Meter units. Examples of standardization and initiatives include, IEC 62056-58, IEC 62056-62 data exchange for meter reading, tariff and load control \cite{Standard_1}  \cite{Standard_2}, Device Language Message Specification DLMS/COSEM protocol \cite{Standard_3}, and communication system for meters and remote reading of meters \cite{Standard_5}.
	
	The standardizations have opened the door for a better system integration of Smart Metering infrastructures for applications; Advanced Metering Infrastructures (AMI), Demand Response and Home Energy Management Systems \cite{SG_Application_DR}. In addition, it has enabled Smart Metering Infrastructures to support ubiquitous communication technologies that can assist continual and remote  monitoring of power system components and greatly improve visibility and optimal control of the power system for power distribution supervisory control and automation systems \cite{SG_Application_SCADA}.
	
	Regarding monitoring and automation of electric distribution system, \cite{DSO_Moni_1} presents a mechanism that utilizes smart metering infrastructure using GSM technology. The developed model is tested using two-way communication for remote control of supply and outage management along with automatic switching-off on the occurrence of fault. 
	
	\cite{DSO_Moni_2},  \cite{DSO_Moni_3} propose detailed examination of active and non-active power components by making use of modern smart meters for studying power quality, load monitoring and active power factor correction. The papers provide methodologies for load characterization, which could possibly benefit consumers and power utilities by improving the power quality. They further tend to show how important it is to analyse non active power components. 
	
	An advanced metering infrastructure to monitor supply parameters, over voltage and over current alerts and real time billing using ADE7758 IC and ATmega 16 microcontroller is proposed by  \cite{DSO_Moni_4}. A significant amount of studies have been done regarding the usage of AMI for energy billing, prepaid metering, remote connection and disconnection of supply to enhance reduction of man power which in-turn leads to less costly systems \cite{DSO_Moni_5} \cite{DSO_Moni_6} \cite{DSO_Moni_7}.
	
	Adaptivity of network QoS and information access configurations is studied in great detail in the context of SmartC2Net project \cite{DSO_Moni_8}, with primary goal of enabling smart grid operations over imperfect, heterogeneous general purpose networks.  \cite{DSO_Moni_9}  \cite{DSO_Moni_10} studied Information-Quality based LV grid monitoring framework and its application to power quality control while  \cite{DSO_Moni_11} discusses utilizing network QoS for dependability of adaptive Smart Grid control  with focus on the influence of imperfect network conditions on smart grid controllers, and how this can be counteracted by utilizing Quality of Service (QoS) information from the communication network. 
	
	The approach proposed in this paper for monitoring of the electrical grid by adaptive data aggregation methods, is focusing on using Smart Metering infrastructures with network level adaptation of the data collection mechanism. The method is unique compared to the above studies in that experimental field tests are used and validated by real-time hardware in the loop platform of Smart Energy Systems Laboratory at AAU \cite{SES_Lab}. Currently most of the smart meters deployed in Denmark are used for meter reading purposes with slow update rates, therefore looking in to aspect of smart meter enabled monitoring of the distribution gird by using adaptive data collection mechanisms would be the main contribution of this work.
	



	\begin{figure*}[]
		\centering
		\includegraphics[width=6.16in,height=3.5in,]{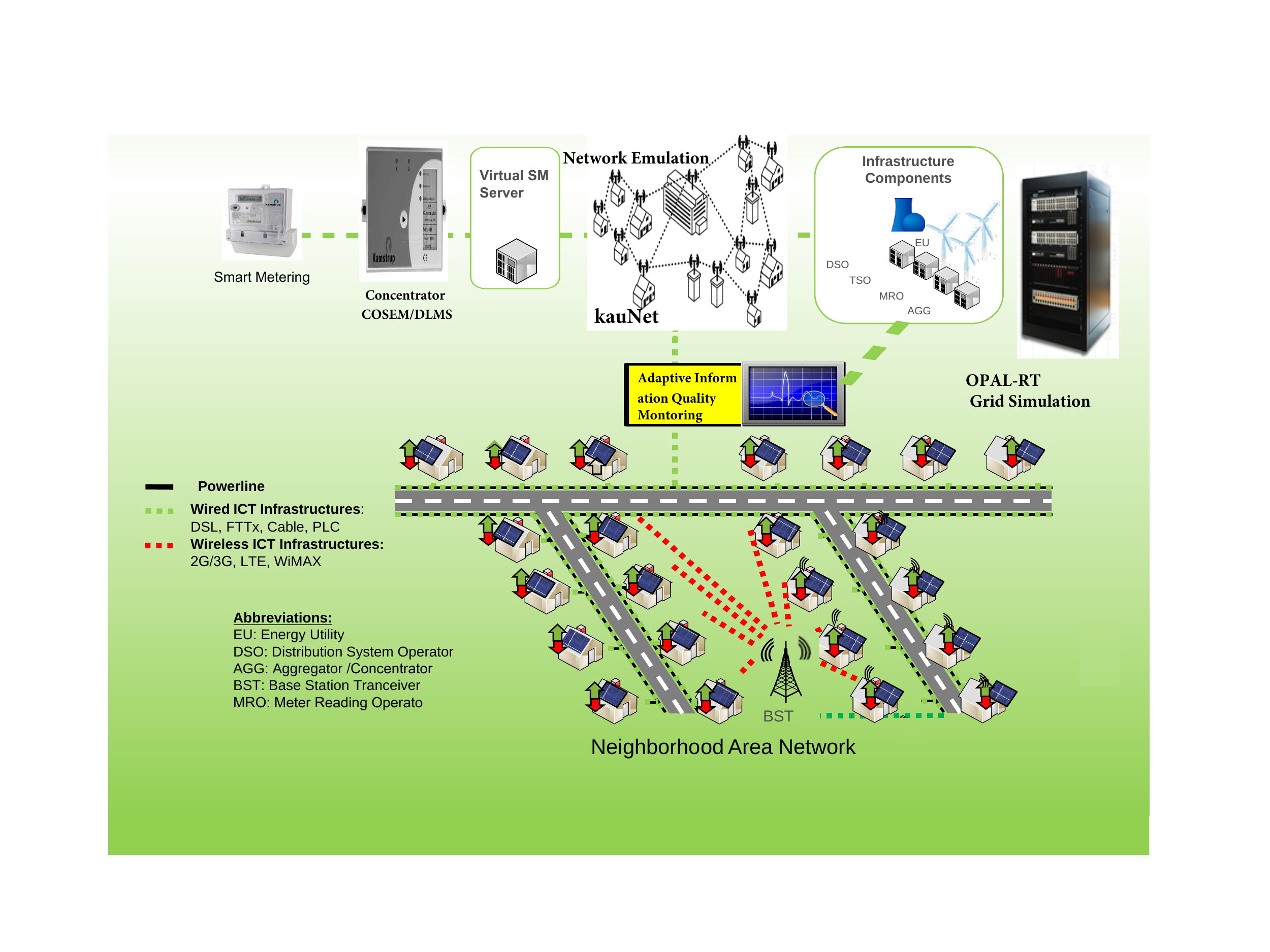}
		\caption{Envisioned Smart Metering Evaluation Platform with Network Emulation using KauNet, Real-Time hardware-in-the-loop (HIL) grid modeling with OPAL-RT, Smart Meters, Industrial Concentrator and Adaptive Information Quality Monitoring. \label{fig:PRESC}} 
	\end{figure*}

\section{Grid Monitoring Use-cases}

The growing integration of renewable energy sources in low voltage grid has brought new challenges in ensuring the power quality due to the unpredictability of these systems. For higher layer controllers, it is evident that an easier and frequent access to the state of these systems is greatly beneficial. After consultation with Thymors Energi, a Danish DSO part of the RemoteGRID project, adaptive data collection  mechanisms will be studied for the following use-cases, with main focus on investigating resilience and dependability of smart data collection mechanisms:

\subsection{Grid Fault Detection and Localization}

\subsubsection {Cable fault } Distribution cable systems are susceptible to faults created by either direct cut, indirect cut and other factors. Finding location of a cable fault is of vital importance for electric distribution system operators, adaptive smart data collection schemes can be beneficial to quickly identify the occurrence of this faulty scenarios.  The mechanisms can be used to notify the monitoring framework and get a better, quick understanding of the problem at hand.

\subsubsection {Converter faults for grid connected photovoltaic systems}  Grid-connected photovoltaic (PV) systems which convert solar energy into electricity are widely deployed in current distribution systems.  The current PV panels have a relatively low and varying output voltage which makes it beneficial for the installed  converter between PVs and the grid to be equipped with high step-up and versatile control capabilities \cite{Adap_App_2}. The output of these systems is rich in harmonics which affects the power quality of the grid.

\subsection{Real Time Grid State Estimation}

State estimation provides monitoring of the grid for many system optimization functionalities. By using  the estimates of bus voltages, load tap changer can adjust the voltage level with a dynamic voltage optimization to facilitate control of the power flow with real-time voltage and current estimates, identification of branches which exceed the electrical limits can be identified, which makes prediction and elimination of faults achievable. In addition the real-time power output of the renewable energy generations can be used to study the impact  on circuit stability and other integration issues. Previously, real-time state estimation has only been implemented on transmission systems by collecting measurements every few seconds but on our current distribution grid, smart meters can provide the required data to do the same for distribution grids \cite{DSO_Moni_12}.

State estimation of electrical grid is essential part of monitoring schemes. With real time current and voltage estimates, buses that surpass electrical limits can be identified. Hence, It is very important that the data collection mechanism should be able to provide information fulfilling the requirement of the grid estimator. State estimation algorithm is not the main focus of this work but it will be done in parallel  as part of the project which focuses mainly on centralized grid estimation techniques.


\section{Adaptive Smart Data Collection Mechanisms }

A detailed analysis and proposal of adaptive smart data collection methods for smart metering data aggregation platforms will be made:

\subsection {Adaptiveness to Smart Metering Application Needs } 

Smart metering data collection for AMI and monitoring applications use periodic or event driven sampling methods \cite{Adap_App_1}. Periodic data collection is also known as data streaming, where data flows primarily from smart meters to higher level aggregators. In the event driven model, smart meter report events for the monitoring mechanism. Event based communication is less demanding in-terms of amount of communication compared to periodic data collection. For monitoring of electrical grid using Smart Metering Infrastructure,  sometimes the \textit{dynamics of the monitored condition or process can change quickly}. If the data collection mechanism can adapt to the changing dynamics of the condition or process, over-sampling can be minimized which is beneficial for power efficiency and to ensure reliable operation within the communication constraints. For the use-cases described in the previous section, a mechanism will be proposed for adaptive smart data collection schemes for monitoring of the metering infrastructure.


\subsection {Adaptiveness to Changing Communication Network Dynamics}
During the course of the research process, a rigorous study will be performed to look into the \textit{communication needs for protocols, Smart Meters, concentrators and smart data collection scheme}. Adaptive system will be developed and tested to cope with communication needs.



\subsection{Adaptiveness to Security Attacks}
One of the major challenges of the Smart Metering infrastructure is protecting the security of the whole system. Security measures should be implemented in controlling confidentiality, availability and integrity to make sure that unauthorized access to the system is not granted. We plan to look at, \textit{how security measures impacts the adaptive system}, \textit{resilience of the adaptive system to smart meter disconnect attacks} and \textit{anomaly detection mechanisms}. The impact of adding a security layer on the adaptive monitoring framework will be studied with the help of the platform.

\section{Research Process and Evaluation Mechanisms}

\par The envisioned Real-Time Smart Metering platform (Figure 2), HIL framework coupled with communication network emulation module will be used to scope the objectives, design the simulation models and define outcomes of the project through the test cases. Real-Time HIL platform is chosen over simulations because it allows realistic timing to control how long a system would take to perform a task or respond to critical events. It also enables to assure that deadlines can be met within predictable and consistent time-lines. For example, the real time grid estimation technique should be tested if it can provide accurate state estimates using the platform.

	

	\subsection{Real-Time Smart Metering HIL platform}
	
	\par To test real industrial concentrators, Smart Meters with a realistic communication mechanisms, it is evident that real-time Hardware in the loop electrical grid platform coupled with realistic network emulation implementation should be used. This experimental platform provides functionalities to perform complex experiments which would otherwise be impossible to test on a real life grid. It has the following main components: 
 
 \subsubsection{Network Emulation}

		Network emulation is commonly used to evaluate and examine the behavior and performance of
		applications and transport layer protocols. It is advantageous over simulations in that we can use a wide range of real implementations of protocols and applications while testing
		the scenarios.
		
		Prior experience on a similar platform has shown that KauNet communication network emulator is a viable platform (Figure 2). It is chosen because of its capability to perform network modeling with large degree of control and repeatability. It also gives the highly desired functionality of placing deterministic delay and packet loss patterns as well as precise control over bandwidth and delay changes \cite{On_Line} and \cite{Kau_Net}. Network parameter traces from the experimental field test will be used to configure the network emulator.
 \subsubsection{Real-Time HIL Grid Model}

		OPAL-RT (Figure 2) will be used to capture the electrical system from the transmission level (TSO) down to low voltage distribution grids. The power system algorithms produce output conditions that realistically represent conditions in a real grid network, making real-time simulation significant for two reasons; the user is capable of performing HIL testing such as with industrial controllers and realistic communication networks can be incorporated as interface between simulated electrical network and the physical components. The OPAL-RT is able to simulate up to 10000 three-phase buses, and implementation of all models is based on MATLAB/SIMULINK. The grid model similar to the experimental field test used will be modelled by a researcher from Energy and Technology department as part of collaboration with the research project. 
		
 \subsubsection{Virtual Smart Meter}				
		
		It will be developed to mimic multiple real smart meters and by using the standard Device Language Message Specification (DLMS) protocol to exchange information with an industrial concentrator (Figure 2), an industrial data aggregation hardware. For the aforementioned use-case, the concentrator will be used for Smart Metering HIL platform implementation. The virtual smart meter platform enables simulation based evaluation of different control and automation strategies to the Smart Metering System. It is favorable from real Smart Meters in that it is very flexible for expansion which can be used to measure and evaluate scalability of the system.

\subsection{Validation and Evaluation}
 \subsubsection{Pilot Field Test Site}		
 
As part of the research on RemoteGRID project, an area in Northern Denmark with 2000 houses which contains, small residential villages, Agricultural sites, small industries with penetration of PV and Wind power generation units will be used for experimental tests and collection of raw data required for the research work.

	\subsubsection{Validation, Data Logging and Analysis}
	
	\par At this phase, a thorough investigation of the data sets collected from all the test cases will be performed. A validation process will be undertaken to showcase the performance of the proposed adaptation mechanisms and proof-tested on real-time HIL platform by using statistical analysis techniques.

\subsubsection {Impact of Communication Disconnection}This impact of communication disconnection on the monitoring of the electrical grid will be tested thoroughly. Focus is given on recognizing the ways in which information and communication network failures cause a loss of control on the monitoring of the distribution grid.

\section{Summary/Outlook}
This article provides a general overview of a proposal for developing, testing and implementing
mechanisms to adapt data collection mechanisms for monitoring of the distribution grid through the existing Smart Metering infrastructure. The envisioned smart metering platform with network emulator and real time HIL platform, virtual smart meters and industrial concentrator is presented. Three important adaptivity aspects, namely adaptiveness to application needs, adaptiveness to dynamic communication needs and adaptiveness to security attacks is proposed on the paper. It is presented that experimental field tests are also important aspect of the project in order to have practical data sets to be used during the simulation/emulations phases to analyse and evaluate the adaptiveness of the system.



\section*{Acknowledgment}

This work is financially supported by the Danish FORESKEL Energinet program under grant agreement no. 2016-1-12399.



%

\end{document}